 \documentclass[preprint,12pt]{elsarticle}

\usepackage{graphicx}

\usepackage{amssymb}

\journal{Computer Physics Communications}

\newcommand{\Revised}[1]{#1}

\newcommand{\F}{{\sc FiEstAS}}

\newcommand{\dd}{{\rm d}}

\newcommand{\vv}[1]{{\bf #1}}
\newcommand{\xx}{\vv{x}}
\newcommand{\XX}{\vv{X}}
\newcommand{\ff}{\hat{f}}
\newcommand{\fx}{\ff(\xx)}
\newcommand{\hh}{\vv{h}}
\newcommand{\HH}{\vv{H}}

\newcommand{\nn}{N_{\rm node}}
\newcommand{\nnei}{N_{\rm nei}}

\newcommand{\spl}{_{\rm split}}

\begin{document}

\begin{frontmatter}



\title{\Revised{Estimating multidimensional probability fields using the Field Estimator for Arbitrary Spaces (FiEstAS) with applications to Astrophysics}}


\author{Yago Ascasibar}

\address{Universidad Aut\'onoma de Madrid\\ Dpto F\'{i}sica Te\'orica, Campus de Cantoblanco, Madrid E-28049, Spain}
\ead{yago.ascasibar@uam.es}

\begin{abstract}
The Field Estimator for Arbitrary Spaces (\F) computes the continuous probability density field underlying a given discrete data sample \Revised{in multiple, non-commensurate dimensions.
The algorithm works by constructing a metric-independent tessellation of the data space based on a recursive binary splitting.
Individual, data-driven bandwidths are assigned to each point, scaled so that a constant ``mass'' $M_0$ is enclosed.
Kernel density estimation may then be performed for different kernel shapes, and a combination of balloon and sample point estimators is proposed as a compromise between resolution and variance.
A bias correction is evaluated for the particular (yet common) case where the density is computed exactly at the locations of the data points rather than at an uncorrelated set of locations.
By default, the algorithm combines a top-hat kernel with $M_0=2.0$ with the balloon estimator and applies the corresponding bias correction.
These settings are shown to yield reasonable results for a simple test case, a two-dimensional ring, that illustrates the performance for oblique distributions, as well as for a six-dimensional Hernquist sphere, a fairly realistic model of the dynamical structure of stellar bulges in galaxies and dark matter haloes in cosmological N-body simulations.
Results for different parameter settings are discussed in order to provide a guideline to select an optimal configuration in other cases.}
Source code is available upon request.
\end{abstract}


\end{frontmatter}


\section{Introduction}
\label{secIntro}

Given a point process where the $D$-dimensional probability density field $f(\xx)$ is sampled by $N$ random points $\XX_i$, the goal of density estimation is to infer the continuous function $f(\xx)$ from the discrete set of $\XX_i$.
One of the most popular approaches to the problem is kernel density estimation, in which the field is estimated by
\begin{equation}
 \fx = \frac{1}{|\HH|} \sum_{i=1}^{N} K\!\!\left(\ \HH^{-1}(\xx-\XX_i)\ \right)
\end{equation}
where the kernel $K(u)$ is an even function that integrates to unity, and the bandwidth $\HH$ is a $D \times D$ matrix that specifies the scale, shape, and orientation of the kernel.
The choice of this matrix has been thoroughly discussed in different contexts, and extensive reviews exist in the literature \cite[e.g.][]{Silverman86,WandJones95}.

The importance of density estimation cannot be overstressed.
Quite often, one is directly interested in the density itself; the \F\ algorithm was originally developed \cite{AscasibarBinney05} to evaluate the density of particles in the six-dimensional phase space of positions and velocities.
Although the problem has recently arisen considerable interest \cite[e.g.][]{Vogelsberger+08,Wojtak+08,Vass+09,Maciejewski+09}, it is of course only an anecdotical example.
Nevertheless, it illustrates the difficulty of defining a metric (and related concepts, such as neighbourhood) in the general, non-Euclidean case.
Although distances can be trivially defined in both three-dimensional subspaces, it is not clear how positions and velocities should be combined in order to produce a meaningful six-dimensional distance.
It can be shown that a global scaling will only be appropriate for a certain region of the phase space, but not for the whole system \cite[see the discussion in][]{AscasibarBinney05,Maciejewski+09}.
In other words, the metric must adapt to the \emph{local} structure of the data in order to recover the underlying density field.

In terms of applications, density estimation can be helpful in data mining problems.
Unsupervised classification may be performed by identifying independent clusters with local density maxima, with boundaries set by the saddle points.
In supervised classification, one can compute the probability distribution for each group $c$ in the training set, $f_c(\xx)$, from the $N_c$ data points belonging to it.
Applying Bayes' theorem, the probability that a new datum $\xx$ belongs to class $c$ is given by
\begin{equation}
 p(c|\xx) = \frac{\pi_c f_c(\xx)}{\sum_i \pi_i f_i(\xx)}
\end{equation}
where $\pi_c$ denotes the prior probability of each class, and the sum in the denominator runs over all classes.

This work discusses the implementation of kernel smoothing in the Field Estimator for Arbitrary Spaces (\F).
The algorithm is fully described in Section~\ref{secFiEstAS}, and the results of benchmark tests are presented in Section~\ref{secResults}.
The main conclusions are summarized in Section~\ref{secConclusions}.

\section{Description of the algorithm}
\label{secFiEstAS}

\F\ provides, for a given dataset $\left\{\XX_i\right\}_{i=1,N}$ in $D$ dimensions, the value of $f(\xx)$ at any arbitrary point $\xx$.
The algorithm involves the following steps:
\begin{enumerate}
 \item Tesselation of the $D$-dimensional space.
 \item Assignment of bandwidths to every data point.
 \item Estimation of $f(\xx)$.
 \item Bias correction (if necessary).
\end{enumerate}

Each of them is described below, along with the different options and parameters that apply in each case.

\subsection{Tesselation}

The first step of the algorithm is the division of the data space in cells containing exactly one point.
An important issue is the absence of a well-defined metric,
which greatly increases the range of applicability of the method.
Rather than using distances between data points, \F\ recursively divides the space by means of a $k$-d tree, one dimension at a time, until there is only one point per leaf.

There are several criteria to select the dimension to split at each step.
The original version of \F\ \cite{AscasibarBinney05} was fine-tuned to estimate densities in phase space, and it used the information that both the position and velocity subspaces are Euclidean.
Moreover, it was imposed that divisions should take place alternatively in each subspace.
A significant improvement over this scheme, proposed by \cite{SharmaSteinmetz06}, is the selection of the dimension with lower Shannon entropy.
Such a choice results in more divisions along the dimensions that show more structure, and therefore it adapts better to the distribution of the data.
A very similar scheme was implemented in \cite{Ascasibar08} to use \F\ in the context of Monte Carlo numerical integration: when a tree node has to be split, a histogram with $B=1+\sqrt{\nn}$ bins is built for each dimension, from the minimum to the maximum value attained by the corresponding coordinate.
The log-likelihood for the histogram counts $n_{b}$ to arise from a Poissonian distribution is given by
\begin{equation}
 L_d = \ln(\nn!) - \nn\ln(B) - \sum_{b=1}^{B} \ln(n_{bd}!)
\end{equation}
where the indices $1\le d\le D$ and $1\le b\le B$ denote the dimension and the bin number, respectively, $n_{bd}$ is the number of points in each bin, and $\nn$ is the total number of points in the node.
The dimension with smaller $L$ is divided at the point $x\spl=(x_{\rm l}+x_{\rm r})/2$, where $x_{\rm l}$ is the maximum $x$ of all points lying on the ``left'' side ($b\le b\spl$) and $x_{\rm r}$ is the minimum $x$ of the points lying on the ``right'' ($b>b\spl$) side.
The bin $1\le b\spl<B$ is chosen in order that the number of points on each side is as close as possible to $\nn/2$.

A crude estimate of the density can be obtained as the inverse of the cell volume.
As shown in \cite{AscasibarBinney05}, this estimate is very noisy, and it dramatically underestimates the density of particles near the boundary of the system.
This becomes a critical problem in many dimensions, because the fraction of points affected quickly approaches unity as $D$ increases.
A simple correction was applied in \cite{AscasibarBinney05} to data points at the boundary of the hypercubical domain, and a scheme based on the mean interparticle separation was used in \cite{SharmaSteinmetz06} to adjust the shape of every tree node.
In the present version of \F, such a correction is not necessary.

\subsection{Bandwith assignment}

In principle, one should compute the $D(D+1)/2$ independent coefficients of the bandwith matrix $\HH$ that minimize the mean integrated square error.
However, doing that for every single datapoint can be impractical for large samples, and a simpler prescription has been adopted.

First, the bandwidth matrices are constrained to be diagonal.
Although this is far from optimal when the data are distributed obliquely with respect to the coordinate axes \cite[see e.g.][]{WandJones93,DuongHazelton03,DuongHazelton05}, there is a substantial gain in speed, memory consumption, and code simplicity, by reducing the number of free parameters.
This prescription will work well if the field is well sampled, although anisotropic kernels would perform better in oblique regions where the sampling is sparse.

\begin{figure}
\begin{center}
\includegraphics[width=8cm]{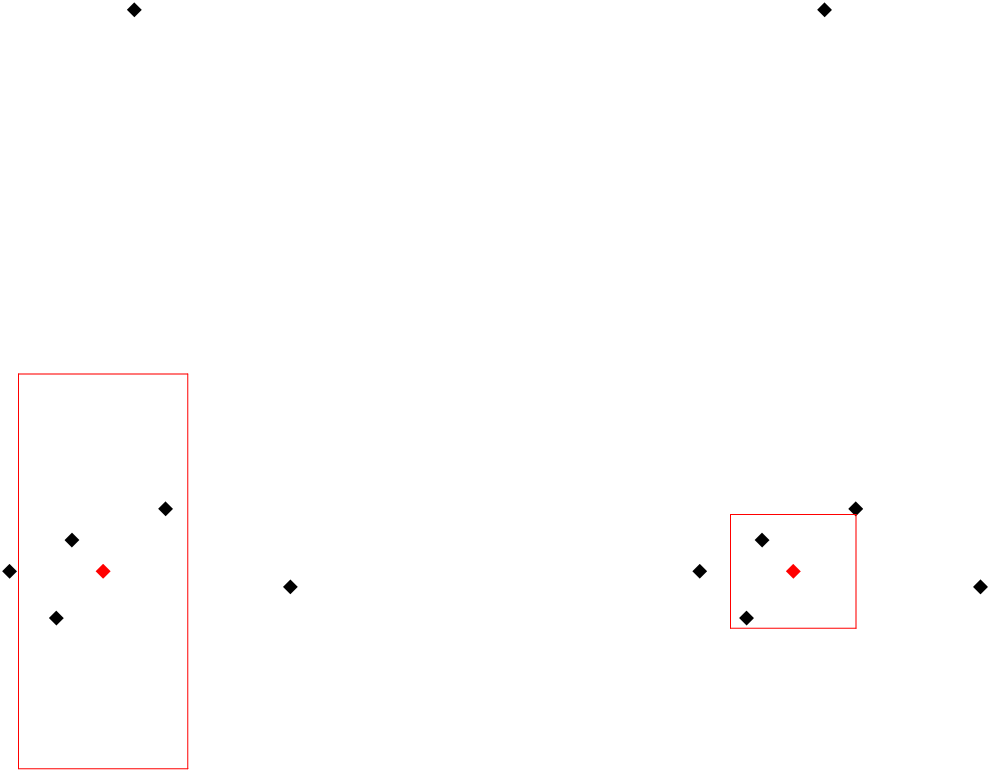}
\end{center}
\caption{
\Revised{Bandwidth assignment for a given particle (plotted in red) in two dimensions.
The box on the left panel represents $\xx\pm\vv{\sigma}$, where $\vv{\sigma}$ is the dispersion vector given by expression~(\ref{eqSigma}).
The bandwidths~(\ref{eqHsmooth}) yield the box $\xx\pm\vv{h}$ shown on the right panel, better adapted to the local distribution of data points.
}}
\label{fig0}
\end{figure}

The relation between the $D$ smoothing lengths $h_d$ of each point is estimated from the \emph{local} dispersion of the data along each axis
\begin{equation}
 \sigma_{d}^2 =
 \sum_{n=1}^{\nnei} X_{nd}^2 -
 \left( \sum_{n=1}^{\nnei} X_{nd} \right)^{\!\!2}
\label{eqSigma}
\end{equation}
where the index $n$ refers to the $\nnei$ neighbours defined by the \F\ tessellation.
The smoothing lengths are then set to
\begin{equation}
 h_{d}^2 =
 \frac{ \sum_{n=1}^{\nnei} w_n X_{nd}^2 }{ \sum_{n=1}^{\nnei} w_n } -
 \left(\frac{  \sum_{n=1}^{\nnei} w_n X_{nd} }{ \sum_{n=1}^{\nnei} w_n }\right)^{\!\!2}
\label{eqHsmooth}
\end{equation}
with weights
\begin{equation}
 w_n = \prod_{d=1}^D \frac{1}{\sigma_d} \exp\left[ -\frac{(X_{nd}-X_{id})^2}{2\sigma_{d}^2} \right]
\end{equation}
This measure is less sensitive to the presence of outliers than the simpler prescription $h_{d} = \sigma_{d}$ \Revised{(see Figure~\ref{fig0})}.

In addition, \F\ offers the possibility of imposing a particular metric to any subspace by specifying a list of dimensions $\{d_l\}_{l=1,L}$ and the relative scale between them $\{s_l\}_{l=1,L}$.
Defining $S=\prod_{l=1}^L s_l$ and $V=\prod_{l=1}^L h_{d_l}$,
\begin{equation}
 h_{d_l} = s_l \frac{V}{S}
\end{equation}
all other dimensions remaining unaltered.
For instance, in phase space one could set dimensions $d_l=\{1,2,3\}$ (positions) to scale as $s_l=\{1.0,1.0,1.0\}$ and then impose the same Euclidean metric to  the velocities, $d_l=\{4,5,6\}$.
The relation between both spaces is not specified, and can vary freely from point to point.

Finally, the overall scale of the bandwidths is set so that the mass contained within the hypercube they define is equal to the user-defined parameter $M_0$.
The value of $M_0$ controls the degree of smoothing, and can be thought of as a constant (not necessarily integer) ``number of neighbours'' of the smoothing kernel.
\Revised{In order to compute it, each data point (of unit mass) is uniformly distributed over its cell, without any boundary correction,
\begin{equation}
 m_i =
 \int_{\XX_i-\hh_i}^{\XX_i+\hh_i} \sum_{j=1}^{N} C_j(\xx)\ \dd^D\xx
\end{equation}
where $C_j(\xx)=1$ if $\xx$ lies inside the $j$-th \F\ cell and 0 otherwise, and the bandwidths are scaled until $m_i=M_0$ within a 10 per cent tolerance.}
This is the only case in which the mass of the data is distributed like in the original implementation of \F.

\subsection{Field estimation}

At this point, it would be possible to estimate the density as
\begin{equation}
 \hat f_K(\xx) =
	\sum_{i=1}^{N} \prod_{d=1}^{D}
	\frac{ 1 }{ h_{id} } K( \frac{ x_d - X_{id} }{ h_{id} } )
\label{eqSamplePoint}
\end{equation}
where we have used a ``product kernel'' $K$.
The current implementation includes top hat, $K(u)=1/2$, triangular-shaped cloud, $K(u)=1-|u|$, and Epanechnikov, $K(u)=\frac{3}{4}(1-u^2)$, kernels, where $-1<u<1$.

Apart from this possibility, \F\ can also combine $\hat f_K(\xx)$ with a top-hat balloon estimator
\begin{equation}
 \hat f_{\rm B}(\xx)
 = \frac{ 1 }{ \prod_{d=1}^{D} 2\hat{h}_{Kd}(\xx) }
	\int_{\xx-\hat\hh_K(\xx)}^{\xx+\hat\hh_K(\xx)} \hat f_K(\xx_0)\ \dd^D\xx_0
\label{eqBalloon}
\end{equation}
based on a local bandwidth
\begin{equation}
 \hat\hh_K(\xx) =
	\frac{1}{\hat f_K(\xx)} \sum_{i=1}^{N} \hh_i \prod_{d=1}^{D}
	\frac{ 1 }{ h_{id} } K( \frac{ x_d - X_{id} }{ h_{id} } )
\end{equation}
interpolated from the individual particle bandwidths $\hh_i$ by using the same kernel as in equation~(\ref{eqSamplePoint}).

\subsection{Bias correction}
\label{secBias}

In many, if not most, practical applications of the algorithm, one is interested in the value of the density field precisely at the locations of the sample points, and only $\hat f_i \equiv \hat f(\XX_i)$ is evaluated.
As discussed in \cite{SharmaSteinmetz06}, a positive bias that depends on the chosen kernel and its bandwidth arises in this particular case because we are not evaluating the density at a completely independent set of locations.
The magnitude of this bias can be easily estimated for a uniform probability distribution by considering the average values of $\hat f_K(\XX_i)$ and $\hat f_{\rm B}(\XX_i)$.
In a \Revised{uniform Poissonian distribution, $f(\xx)=f_0$}, all the smoothing lengths would be given by
\begin{equation}
 M_0 \approx f_0 (2h)^D
\end{equation}
and thus
\begin{equation}
 \langle \hat f_K(\XX_i) \rangle =
	\prod_{d=1}^{D} \frac{ K(0) }{ h } +
	(N-1) \prod_{d=1}^{D} \frac{ \langle K \rangle }{ h }
= \frac{ \left[\,2K(0)\,\right]^D }{M_0} f_0 + \frac{N-1}{N} f_0
\end{equation}
whereas, for the balloon estimator,
\begin{equation}
 \langle \hat f_{\rm B}(\XX_i) \rangle =
	\prod_{d=1}^{D} \frac{ 1 }{ 2h } +
	(N-1) \prod_{d=1}^{D} \frac{ \langle \int_{\XX_i-h}^{\XX_i+h}K \rangle }{ h }
= \frac{ 1 }{M_0} f_0 + \frac{N-1}{N} f_0
\end{equation}

Therefore, assuming $N\gg1$, the algorithm can apply a correction $\hat f_i = \hat f_i^{\rm uncorrected}/(1+b)$ when only the $\hat f_i$ are requested, where $b_K = \left[\,2K(0)\,\right]^D / M_0$ and $b_{\rm B} = 1/M_0$.
\Revised{It is important to bear in mind that this correction factor must \emph{not} be applied in the general case, where the sample and evaluation points do not coincide.
In particular, it should not be confused with the bias arising from the derivatives of $f$ (note that, in fact, the values of $b$ have been derived for a constant density), that has not been accounted for due to the difficulties associated to the estimation of local derivatives.
}

\section{Results}
\label{secResults}

The accuracy of the density reconstruction has been tested in two benchmark cases: a two-dimensional ring and a six-dimensional Hernquist sphere.
We compare the performance of differnet kernels, as well as the scaling with the number $N$ of sample points.
\Revised{Regarding the smoothing parameter, $M_0=2$ arguably represents a reasonable minimum, with smaller values yielding results (bandwidths and densities) that are dominated by the nearest data point.
As will be shown below, increasing this parameter reduces the statistical variance of the estimator at the expense of resolution.
A value $M_0=10$ is considered for reference, but higher values may be suitable depending on the user requirements, especially as the number of dimensions increases.}

\subsection{Two-dimensional ring}

\begin{figure*}
\begin{center}
\includegraphics[width=3cm]{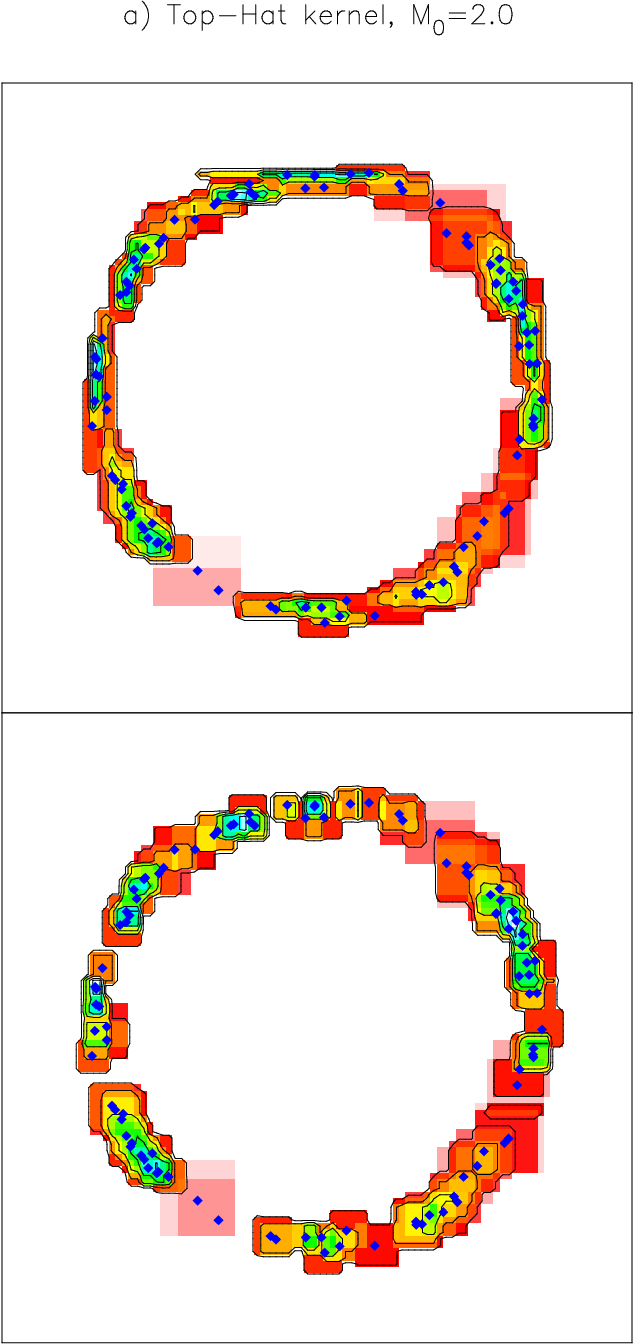}
\includegraphics[width=3cm]{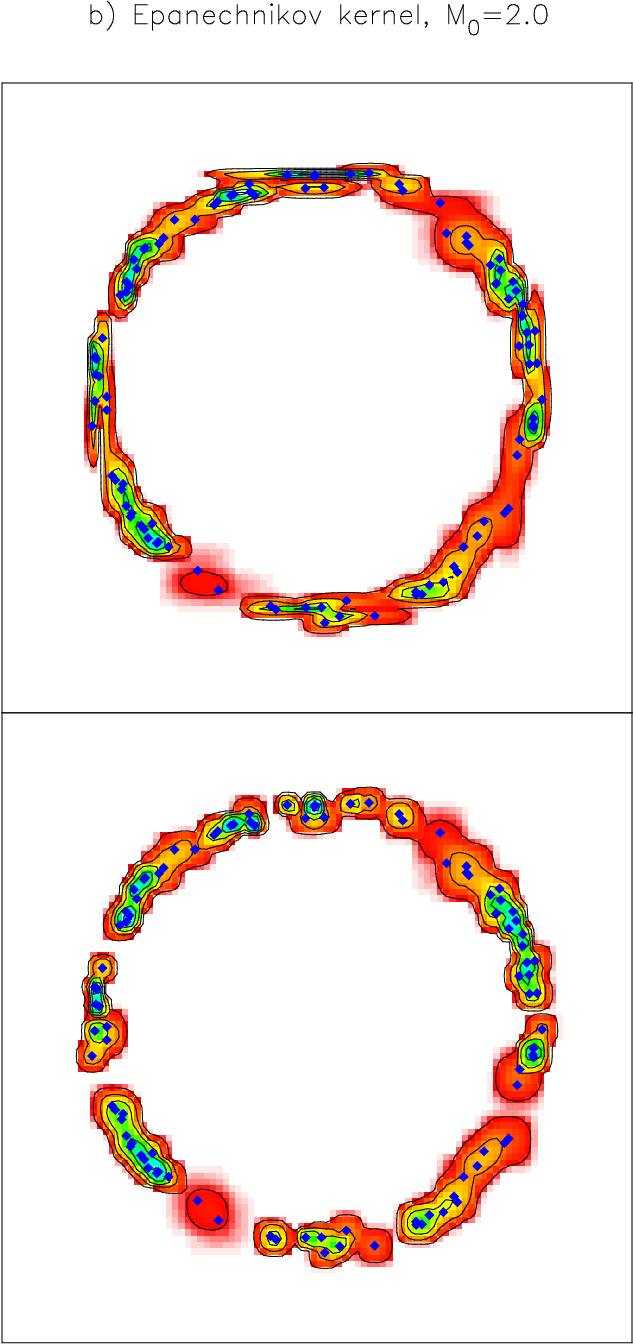}
\includegraphics[width=3cm]{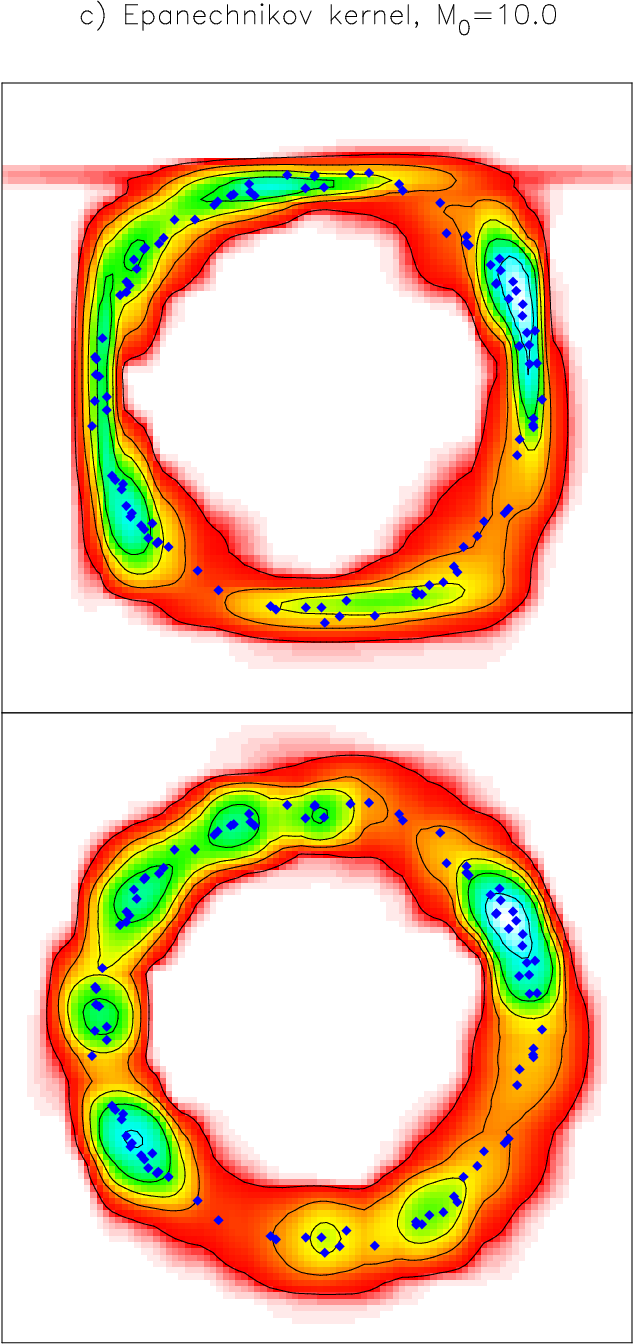}
\includegraphics[width=3cm]{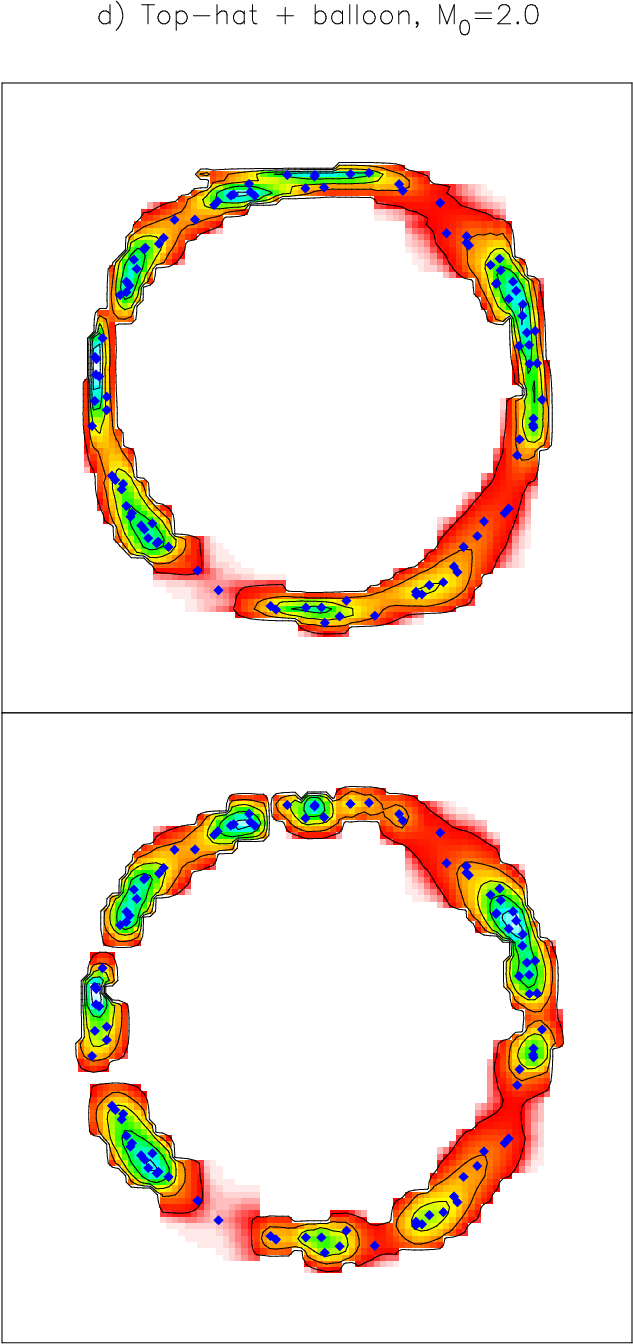}
\end{center}
\caption{
Density field recovered by the \F\ algorithm for a random realization of a two-dimensional ring distribution with 100 sample points (blue squares).
Colours indicate local density, in arbitrary units, and contours enclose 5, 25, 50, 75, and 95 per cent of the mass.
\Revised{Dashed lines indicate the true distribution.}
The metric used on the top panels has not been constrained, whereas an Euclidean metric has been imposed on the bottom panels.
Columns represent the results obtained for: a) top-hat kernel with $M_0=2$. b) Epanechnikov kernel with $M_0=2$. c) Epanechnikov kernel with $M_0=10$. d) \F\ balloon estimator, equation~(\ref{eqBalloon}), combined with a top-hat kernel with $M_0=2$.
}
\label{fig1}
\end{figure*}

The first distribution is a ring in two dimensions with uniform density between an inner and an outer radius of 0.95 and 1.05, respectively, in arbitrary units.
A random realization with 100 sample points is depicted in Figure~\ref{fig1}, together with the density field returned by the \F\ algorithm under different parameter configurations.
In all cases, the shape of the ring is correctly recovered, although some artifacts arise when the cells of the \F\ tessellation become extremely elongated.
Since these artifacts are associated to individual points, they become more evident for large values of $M_0$.
As can be seen in the bottom panels, they are completely absent when a locally Euclidean metric (arguably the most appropriate for this problem, at least globally) is imposed.

The reconstruction obtained by the top-hat kernel has the obvious drawback of the sharp square edges, and the results obtained with the triangular-shaped cloud (not shown) or the Epanechnikov kernel are much more satisfactory in that sense.
For $N=100$, the Epanechnikov kernel with $M_0=10$ tends to severely oversmooth the density distribution.
When the metric is constrained to be locally Euclidean ($h_{\rm x}=h_{\rm y}$ at every point), the width of the ring is systematically overestimated, but the recovered shape is perfectly circular.
For the unrestricted metric, the density distribution is deformed into a slightly square shape aligned with the coordinate axes.
This is due to the \Revised{combined effect of the hypercubical \F\ tessellation (see \cite{Maciejewski+09} for a comparison of different schemes) and the diagonal bandwith matrix.
As a result, kernel shapes in ``horizontal'' or ``vertical'' regions tend to be more elongated, whereas $h_{\rm x} \sim h_{\rm y}$ in the ``diagonal'' regions, causing} the ``diamond'' and ``square'' shapes observed for the inner and outer boundaries of the distribution.
As stated above, it is in these oblique regions, poorly sampled within a smoothing volume, where an anisotropic kernel would certainly provide a significant advantage.
Finally, combining a top-hat kernel with $M_0=2$ with the balloon estimator~(\ref{eqBalloon}) yields a density field that is bracketed by the results of the Epanechnikov kernel with $M_0=2$ and $M_0=10$.

\begin{figure*}
\includegraphics[width=\textwidth]{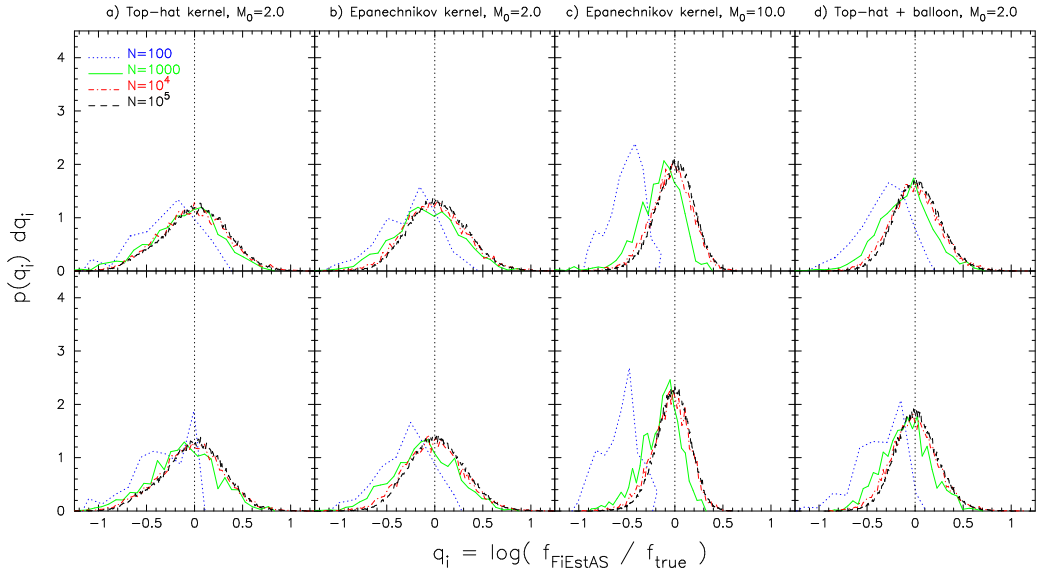}
\caption{
Probability distribution of the variable $q_i=\log\frac{\hat f(\XX_i)}{f(\XX_i)}$ for random realizations of the two-dimensional ring distribution with $N=100$, $1000$, $10^4$ and $10^5$ sample points.
Columns represent different estimators, and an Euclidean metric has been imposed on the bottom panels.
}
\label{fig2}
\end{figure*}

More quantitatively, the probability distribution of the variable $q_i=\log\frac{\hat f(\XX_i)}{f(\XX_i)}$ is shown in Figure~\ref{fig2} for several values of the number $N$ of sample points between $N=100$ and $N=10^5$.
The bias $\langle q_i \rangle$ and the variance $\sqrt{ \langle q_i^2 \rangle - \langle q_i \rangle^2 }$ of each estimator are quoted in Table~\ref{tab2}.
Since the density could already be properly reconstructed with $N\sim 1000$ points, the probability distribution of $q_i$ for this two-dimensional problem does not change much with $N$, with the exception of the oversmoothing shown by all estimators for $N=100$.
The bias correction was of the order of $20-50$ per cent ($0.09-0.18$ dex) in all cases but the Epanechnikov kernel with $M_0=2$, for which it was about a factor of two.
The variance also depends on the choice of a specific kernel and smoothing parameter $M_0$, ranging from $\sim60$ percent in the Epanechnikov kernel with $M_0=10$ to more than a factor of two for the top-hat kernel.
It may be argued, though, that some of this dispersion is indeed physical, in the sense that it reflects the Poisson fluctuations inherent to the random realization of the ideal uniform distribution.
In other words, there really are several clumps in the point distribution, and they are clearly visible in Figure~\ref{fig1}.
\Revised{If one is interested in the actual physical density of these regions, its value should be higher than in those others that happen to contain less points.
If, on the other hand, one is interested in the probability density field from which the sample was drawn, some statistical criterion has to be devised in order to test whether the fluctuations correspond to real variations of the field or are simply due to Poisson noise.}

\begin{table*}
\begin{center}
\begin{tabular}{|ccccc|}
\hline
$N$ & Top-hat & Epanechnikov & Epa., $M_0=10$ & Top-hat+balloon \\ \hline
 $100$ & $-0.28 \pm 0.33 $ & $-0.27 \pm 0.31 $ & $-0.50 \pm 0.18 $ & $-0.32 \pm 0.26 $ \\
$1000$ & $-0.10 \pm 0.38 $ & $-0.09 \pm 0.35 $ & $-0.17 \pm 0.24 $ & $-0.11 \pm 0.29 $ \\
$10^4$ & $-0.03 \pm 0.36 $ & $-0.00 \pm 0.32 $ & $-0.04 \pm 0.22 $ & $-0.01 \pm 0.26 $ \\
$10^5$ & $-0.00 \pm 0.34 $ & $~~0.03 \pm 0.30 $ & $-0.01 \pm 0.21 $ & $~~0.02 \pm 0.24 $ \\ \hline
 $100$ & $-0.33 \pm 0.30 $ & $-0.30 \pm 0.29 $ & $-0.57 \pm 0.19 $ & $-0.36 \pm 0.24 $ \\
$1000$ & $-0.12 \pm 0.35 $ & $-0.11 \pm 0.34 $ & $-0.15 \pm 0.21 $ & $-0.09 \pm 0.25 $ \\
$10^4$ & $-0.04 \pm 0.34 $ & $-0.01 \pm 0.31 $ & $-0.05 \pm 0.20 $ & $-0.01 \pm 0.25 $ \\
$10^5$ & $-0.02 \pm 0.32 $ & $~~0.02 \pm 0.29 $ & $-0.03 \pm 0.18 $ & $~~0.01 \pm 0.23 $ \\ \hline
\end{tabular}
\end{center}
\caption{
Average value $\langle q_i \rangle$ and dispersion $\sqrt{ \langle q_i^2 \rangle - \langle q_i \rangle^2 }$ of the variable $q_i=\log\frac{\hat f(\XX_i)}{f(\XX_i)}$ for the two-dimensional ring distribution.
Columns show the number of sample points and the results of each estimator.
Top and bottom rows correspond to the unrestricted and Euclidean metrics, respectively.
}
\label{tab2}
\end{table*}

\subsection{Hernquist sphere}

The performance of the algorithm has also been tested by recovering the density of a six-dimensional Hernquist sphere \cite{Hernquist90}.
This distribution is often used to model the central bulges of galaxies, as well as their dark matter haloes.
The density of particles in the phase space of three-dimensional positions~$\vv{r}$ and velocities~$\vv{v}$ can be written as
\begin{equation}
 f(\vv{r},\vv{v}) = 
\frac{M/a^3}{4\pi^3\left(2GM/a\right)^{3/2}}
\frac{ 3\sin^{-1}\sqrt{\epsilon}+\sqrt{\epsilon(1-\epsilon)}(1-2\epsilon)(8\epsilon^2-8\epsilon-3) }
{ \left(1-\epsilon\right)^{5/2} }
\end{equation}
in terms of the dimensionless specific binding energy of the particle
\begin{equation}
 \epsilon = \frac{1}{1+r/a}-\frac{v^2}{2GM/a}
\end{equation}
and the total mass $M$ and characteristic radius $a$ of the system.
The generation of a random realization of this distribution is described in \cite{AscasibarBinney05}.

\begin{figure*}
\begin{center}
\includegraphics[width=\textwidth]{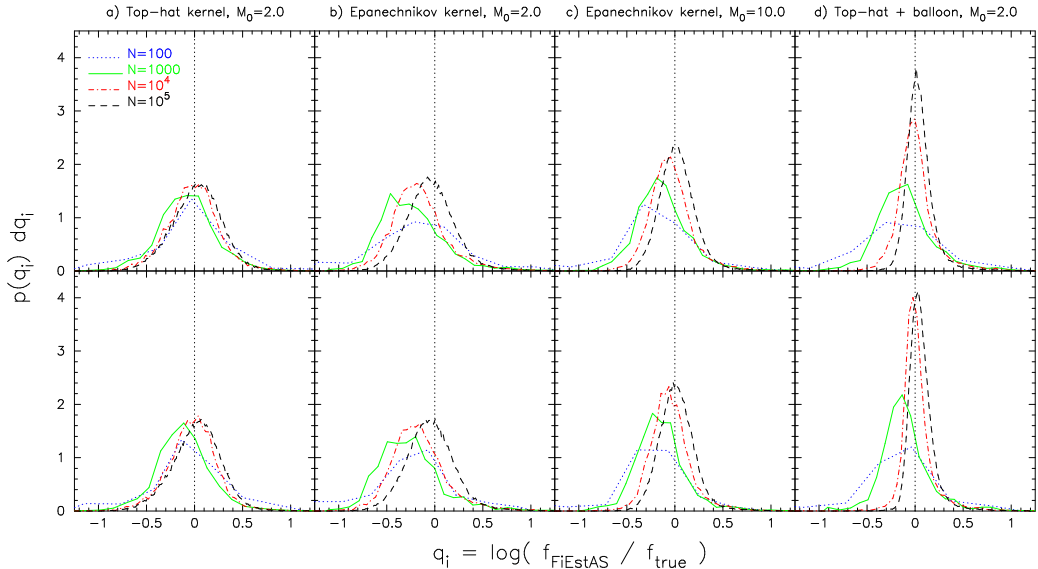}
\end{center}
\caption{
Probability distribution of the variable $q_i=\log\frac{\hat f(\XX_i)}{f(\XX_i)}$ for the six-dimensional Hernquist sphere.
On the bottom panels, a three-dimensional Euclidean metric has been imposed locally to both the position and velocity subspaces.
}
\label{fig3}
\end{figure*}

\begin{table*}
\begin{center}
\begin{tabular}{|ccccc|}
\hline
$N$ & Top-hat & Epanechnikov & Epa., $M_0=10$ & Top-hat+balloon \\ \hline
 $100$ & $-0.07 \pm 0.46 $ & $-0.16 \pm 0.49 $ & $-0.25 \pm 0.49 $ & $-0.21 \pm 0.56 $ \\
$1000$ & $-0.08 \pm 0.31 $ & $-0.24 \pm 0.34 $ & $-0.13 \pm 0.29 $ & $-0.11 \pm 0.31 $ \\
$10^4$ & $-0.01 \pm 0.28 $ & $-0.16 \pm 0.30 $ & $-0.04 \pm 0.24 $ & $~~0.01 \pm 0.22 $ \\
$10^5$ & $~~0.03 \pm 0.26 $ & $-0.04 \pm 0.26 $ & $~~0.03 \pm 0.20 $ & $~~0.05 \pm 0.16 $ \\ \hline
 $100$ & $-0.10 \pm 0.44 $ & $-0.21 \pm 0.49 $ & $-0.28 \pm 0.48 $ & $-0.24 \pm 0.52 $ \\
$1000$ & $-0.12 \pm 0.29 $ & $-0.26 \pm 0.33 $ & $-0.15 \pm 0.28 $ & $-0.10 \pm 0.27 $ \\
$10^4$ & $-0.02 \pm 0.26 $ & $-0.17 \pm 0.30 $ & $-0.05 \pm 0.23 $ & $~~0.02 \pm 0.19 $ \\
$10^5$ & $~~0.02 \pm 0.26 $ & $-0.04 \pm 0.26 $ & $~~0.02 \pm 0.20 $ & $~~0.06 \pm 0.14 $ \\ \hline
\end{tabular}
\end{center}
\caption{
Average value $\langle q_i \rangle$ and dispersion $\sqrt{ \langle q_i^2 \rangle - \langle q_i \rangle^2 }$ of the variable $q_i=\log\frac{\hat f(\XX_i)}{f(\XX_i)}$ for the six-dimensional Hernquist sphere.
}
\label{tab3}
\end{table*}

Results obtained for different values of $N$ are displayed in Figure~\ref{fig3} and Table~\ref{tab3}.
Overall, they are qualitatively similar to the example discussed in the previous section, with only minor differences due to the higher dimensionality of the problem and the very inhomogeneous nature of the Hernquist density distribution.
In particular, the bias correction is much more important in six dimensions, reaching values as high as a factor of $\sim6.7$ for the Epanechnikov kernel with $M_0=2$.
Moreover, many more points are necessary in order to achieve an adequate sampling, and a clear evolution with $N$ is now evident in the probability distribution of $q_i$.
The negative bias observed at low $N$ is mostly due to oversmoothing of the central regions, which contain the majority of the particles.
The Hernquist distribution becomes optimally resolved for $N=10^4-10^5$: the sampling within a smoothing volume becomes close to Poissonian, and the probability distribution of $q_i$ approaches the asymptotic for the chosen kernel.
As in the two-dimensional ring, the specification of a metric based on external knowledge of the problem (in this case, $h_1=h_2=h_3$ and $h_4=h_5=h_6$) affects the results only mildly.

\section{Conclusions}
\label{secConclusions}

Kernel density estimation has been implemented within the Field Estimator for Arbitrary Spaces (\F) algorithm, using different kernels and opening the possibility of combining sample point and balloon estimators.
The only free parameters are the specific form of the kernel function (top-hat, triangular-shaped cloud and Epanechnikov kernels are provided by default) and the smoothing parameter $M_0$.
The bandwidth matrix, constrained to be diagonal, is automatically computed for every point.
Additional constraints can be imposed by the user, but the test cases considered do not suggest that this results in a significant advantage.
\Revised{In fact, it has already been established for a wide range of cases \cite[see e.g.][]{WandJones93} that independent bandwidths (arbitrary metric) do not lose power against the Euclidean metric, even if the latter is true.}
A bias correction must be applied when one is only interested in the values of the density field exactly at the sample points $\XX_i$.
The magnitude of this correction depends on the details of the kernel, but it is already significant at $D=2$ and tends to increase with dimensionality.

The optimal choice of kernel and smoothing parameter are, of course, problem-dependent.
Based on the results presented in the previous section, the combination of a top-hat kernel with $M_0=2$ with the balloon estimator given by equation~(\ref{eqBalloon}) seems to yield a reasonable compromise between accuracy (low dispersion) and resolution (small number of points required) for any number $D$ of dimensions.
This, however, may not hold in the general case, and the user is encouraged to experiment with different options.
\Revised{In particular, smaller values of the smoothing parameter $M_0$ are unlikely to provide useful results, but larger bandwidths may be helpful in order to reduce the statistical noise of the estimator at the expense of losing information about the small-scale structure of the data.
The kernel shape has a much milder effect, but in some cases (e.g. if exact mass conservation is required), a sample point may be preferable to a balloon estimator.
In this case, the Epanechnikov kernel is optimal for an $L_2$ loss criterion with fixed bandwidths \cite{WandJones95}, and this would be, in principle, the recommended choice.}

\section*{Acknowledgments}

Financial support for this work has been provided by the Spanish \emph{Ministerio de Educaci\'on y Ciencia} (project AYA2007-67965-C03-03) and the European Science Foundation (ESF) for the activity entitled ``Computational Astrophysics and Cosmology'' (reference ASTROSIM 2027).





 \bibliographystyle{elsarticle-num}
 \bibliography{bibliography}


\end{document}